\documentclass[%
 reprint,
 amsmath,amssymb,
 aps,
 prl,
]{revtex4-1}
\usepackage{slashed}
\usepackage{graphicx}
\usepackage{dcolumn}
\usepackage{bm}

\begin{document}

\noindent \\
\begin{flushright} 
arXiv:1812.XXXX [hep-th] \\ 
CERN-TH-2018-271 \\
December 2018 \\
\end{flushright}

\title{Skewed Sudakov Regime, Harmonic Numbers, and Multiple Polylogarithms}

\author{Victor T. Kim$^{\it ab}$, 
Victor A. Matveev$^{\it cd}$ and Grigorii B. Pivovarov$^{\it d}$}
 \affiliation{$^{\it a}$ Petersburg Nuclear Physics Institute NRC KI, Gatchina, Russia \\
 $^{\it b}$ St. Petersburg Polytechnic University, St. Petersburg, Russia \\
 $^{\it c}$  Joint Institute for Nuclear Research, Dubna, Russia \\
 $^{\it d}$ Institute for Nuclear Research RAS, Moscow, Russia \\
 }

\date{\today}

\begin{abstract}
On the example of massless QED we study an asymptotic of the vertex when
only one of the two virtualities of the external fermions is sent to zero. We
call this regime the skewed Sudakov regime. First, we show that the asymptotic
is described with a single form factor, for which we derive a linear evolution 
equation. 
The linear operator involved in this equation has a discrete spectrum. Its 
eigenfunctions and eigenvalues are found. The spectrum is a shifted sequence of 
harmonic numbers. With the spectrum found, we represent the expansion of the 
asymptotic in the fine structure constant in terms of multiple polylogarithms. Using this 
representation, the exponentiation of the doubly logarithmic corrections of the Sudakov
form factor is recovered. It is pointed out that the form factor of the skewed Sudakov
regime is growing with the virtuality of a fermion decreasing at a fixed virtuality of 
another fermion.
\end{abstract}

\pacs{Valid PACS appear here}
\maketitle


\section{Introduction}

We start with recalling the classical result of Ref.~\cite{Sudakov:1954sw} introducing along the
way our notations. Let $\Gamma_\mu(p,p')$ be a connected three-point amplitude of massless QED
renormalized with minimal subtractions. Here $p$ and $p'$ are the momenta of the incoming and 
outgoing
fermion respectively. We use namely a \textit{connected} amplitude because it will be technically 
important for our
consideration to include self-energy corrections to one of the external fermion legs of $\Gamma_\mu(p,p')$. 
We also exclude
the overall factor $ie$ from the definition of $\Gamma_\mu(p,p')$. Due to this, the expansion of 
$\Gamma_\mu(p,p')$ in 
the coupling 
starts with Dirac gamma matrix,
$\Gamma_\mu(p,p')=\gamma_\mu+\dots.$

To avoid infrared problems, one considers $\Gamma_\mu(p,p')$ at negative virtualities of the 
external fermions, $p^2<0,p'^2<0$, and of the external photon, $(p'-p)^2<0.$ The results of
\cite{Sudakov:1954sw} are for the kinematics when, on top of the above restrictions,
the Minkowski product of the fermion momenta satisfies the inequality preventing Wick rotation, 
$(pp')^2>p^2p'^2$.
These conditions combined imply that if $|p^2|$ and $|p'^2|$ are small with respect to 
$|(p-p')^2|$, $pp'>0$.
We will use two positive variables to describe this kinematics:
\begin{equation}
    \label{xy}
    x=\frac{-p^2}{2pp'}>0,\,y=\frac{-p'^2}{2pp'}>0.
\end{equation}

With the notations introduced, Sudakov result reads
\begin{equation}
    \label{Sudakov}
    \Gamma_\mu(p,p')\approx\gamma_\mu\exp\big(-\frac{\alpha}{2\pi}\log(x)\log(y)\big),
\end{equation}
where the approximate equality holds when both $|p^2|$ and $|p'^2|$ are small with respect to $|(p'-p)^2|$, 
and their ratio
is of order unity. Also, $\alpha$ is the fine structure constant normalized at any of the small 
virtualities. 

In this paper we derive a generalization of the approximation (\ref{Sudakov}) valid in a wider region
where the magnitude of only one of the two fermion virtualities is small with respect to the 
magnitude of the virtuality of the photon. Because
the symmetry between the two fermion virtualities present in the Sudakov regime is lost for this
generalization, we call this generalized regime  the \textit{skewed} Sudakov regime. For definiteness, we
consider the kinematics with $|p^2|$ small with respect to the magnitude of the photon virtuality, and no 
restrictions on $|p'^2|$ is present. All the considerations can be repeated with the obvious 
changes for the case when $|p'^2|$ is small instead of $|p^2|.$

After the external fermion with small virtuality magnitude is specified, we can also specify
the external fermion leg of $\Gamma_\mu(p,p')$ with self-energy corrections included---it
is the leg of small virtuality. Also, as before, the fine structure constant in the 
subsequent formulas is normalized at the small virtuality. 

To write down the generalization of the approximation (\ref{Sudakov}) we introduce the 
following variables:
\begin{equation}
    \label{tz}
    t=-\frac{\alpha\log(x)}{2\pi}>0,\,z=\frac{1}{1+y}.
\end{equation}
Notice that $0<z<1$, and, in the Sudakov regime $z\to 1$ from the left. With these variables the 
approxiamtion valid in the skewed 
Sudakov regime reads
\begin{equation}
    \label{sS}
    \Gamma_\mu(p,p')\approx\gamma_\mu F(t,z)+
    \frac{p_\mu\slashed{p}'}{pp'}\big[e^{t/2}-F(t,z)\big],
\end{equation}
where the form factor $F(t,z)$ is as follows:
\begin{equation}
    \label{F}
    F(t,z)=e^{3t/2}\sum_{n=0}^\infty e^{-tH_{n+1}}z^n(1-z).
\end{equation}
Here $H_{n+1}$ are the harmonic numbers,
\begin{equation}
    \label{H}
    H_k=\sum_{i=1}^k\frac{1}{i}.
\end{equation}
We point out that the series in the right hand side of Eq. (\ref{F}) converges at $|z|<1$, and to 
recover the approximation (\ref{Sudakov}) one has to send $z\to 1$, where the convergence fails.

To overcome this difficulty, we derive from Eq. (\ref{F}) the following representation:
\begin{equation}
    \label{L}
    F(t,z)=e^{3t/2}\big[1+
    \frac{1}{z}\sum_{\underline{s}}\frac{(-t)^{|\underline{s}|}}{\underline{s}!}
    Li_{\underline{s}}(z)\big],
\end{equation}
where the sum runs over strings of positive integers of arbitrary depth
$d$, $\underline{s}=(s_1,s_2,\dots,s_d)$; $|\underline{s}|$ denotes the weight of the string,
$$|\underline{s}|=\sum_{i=1}^d s_i,$$
and the factorial of the string is the product of the factorials,
$$\underline{s}!=\prod_{i=1}^d s_i!$$
The key ingredient of the representation (\ref{L}) is the so called multiple 
polylogarithm $Li_{\underline{s}}(z).$ 
See,  e. g., Ref.~\cite{Waldschmidt} for an introductory exposition of these functions, and Refs.~\cite{Weinzierl:2007cx,Bogner:2012dn} for their recent applications to Feynman integrals.

Using the properties of multiple polylogarithms discussed in Ref.~\cite{Waldschmidt} one can single out the terms of
the sum of Eq. (\ref{L}) most singular in the limit $z\to 1$. By ``most singular'' we mean that each power of $t$
is compensated by a power of $\log(1-z)$. These are the terms with the strings $\underline{s}$ consisting of units,
$\underline{s}_d=(1,\dots,1)$, with the unit repeated $d=|\underline{s}|$ times. Explicitly,
\begin{equation}
    \label{L-s-d}
    Li_{\underline{s}_d}(z)=(-\log(1-z))^d/d!\approx(-\log(y))^d/d!
\end{equation}
Using this and taking into account that $\slashed{p'}$ is small in the Sudakov regime, one reproduces
approximation (\ref{Sudakov}) from approximation (\ref{sS}).

On the other hand, at fixed $z<1$ and large $t$ one picks up the infinitely growing term in Eq. (\ref{F}) and
obtains the following approximation valid at $x\to +0$ and $y>0$ fixed:
\begin{equation}
    \label{y}
    \Gamma_\mu(p,p')\approx\frac{y}{(1+y)x^{\alpha/(4\pi)}}\big[\gamma_\mu+2\frac{p_\mu\slashed{p}'}{-p'^2}\big].
\end{equation}

We conclude that Eqs. (\ref{sS}) and (\ref{F}) give a unified description for qualitatively different 
asymptotics of the vertex, one of which coincides with the known
Sudakov asymptotic.

The validity of approximation (\ref{sS}), (\ref{F}), and the representation (\ref{L}) are the main results 
of the paper. In the rest of the paper we sketch the way these results are derived.

\section{Schwinger-Dyson Equation and the Inclination Variable}

The vertex $\Gamma_\mu(p,p')$ coincides with its bare counterpart because its renormalization constant
$Z_\Gamma=1$ due to Ward identities between the renormalization constants of QED. With this observation
one can derive the following Schwinger-Dyson equation for the vertex:
\begin{widetext}
\begin{equation}
    \label{SD}
   \Gamma_\mu(p,p')=\gamma_\mu
-\alpha\int\frac{2d^4k}{(2\pi)^3}\Gamma_{\mu\lambda}(k,p';p-k)
   \frac{\slashed{k}\gamma_\rho}{k^2}\big(iD^{\lambda\rho}(p-k)\big).
\end{equation}
\end{widetext}
Here $\Gamma_{\mu\nu}(k,p';p-k)$ is a renormalized connected four point amplitude with $k$
and $p'$ denoting the momenta of incoming and outgoing fermion, and $p-k$ is the incoming momentum of
a photon with Lorentz index $\nu$. Self-energy corrections are included in its $k$-leg, and the amplitude 
is divided by $(ie)^2$. Its expansion in the coupling starts as follows:
\begin{equation}
    \label{mu-nu-exp}
    \Gamma_{\mu\nu}(k,p';l)=\gamma_\nu\frac{i}{\slashed{p}'-\slashed{l}}\gamma_\mu+
    \gamma_\mu\frac{i}{\slashed{k}+\slashed{l}}\gamma_\nu+\dots.
\end{equation}

Another ingredient in Eq. (\ref{SD}) is the full photon propagator $D^{\nu\rho}(p-k)$. 
We use an arbitrary covariant gauge to define it.

Notice that the integration in the right hand side of Eq. (\ref{SD}) is ultraviolet (UV) finite. 
This is because the UV divergences 
originating from various contributions to the four point amplitude under the integral 
cancel against each other due to
gauge invariance. For example, in the one loop approximation, the first term in the 
right hand side of Eq. (\ref{mu-nu-exp}) produces an 1PI contribution to the vertex, 
while the second, the self energy correction to the $p$-leg, and their sum is UV finite. 
We conclude that there is no need to regularize the integration in any way.

Next we change the integration variables. A fist step in this direction is to decompose $k$ into 
transverse and longitudinal parts, the longitudinal part is a
linear combination of the external momenta $p$ and $p'$, and
the transverse part has zero Minkowski products with them:
\begin{align}
    \label{decomp}
    &k=k_\perp+k_\parallel, \nonumber\\
    &k_\parallel=\alpha p+\beta p',
    pk_\perp =0,p'k_\perp =0.
\end{align}
Because of the above condition $(pp')^2>p^2p'^2$,
$k_\perp$ is necessarily euclidean, and we represent
the measure of integration from Eq. (\ref{SD}) in the polar coordinates:
\begin{equation}
    \label{polar}
    2d^4k\to d^2k_\parallel \big(\theta(\phi)
    \theta(2\pi-\phi)d\phi\big)dk_\perp^2
    \theta(-k^2_\perp),
\end{equation}
where the transverse components are 
\begin{equation}
    \label{components}
    k_1=\cos(\phi)\sqrt{-k_\perp^2}, k_2=\sin(\phi)\sqrt{-k_\perp^2}.
\end{equation}

Next we observe that if there were no restriction on the sign of $k_\perp^2$
in the last factor of Eq. (\ref{polar}), the integral in Eq. (\ref{SD}) would vanish.
This happens because of the analytic properties of the integrand. In the one loop 
approximation this is evident from the explicit form of Eq. (\ref{mu-nu-exp}), and 
is less evident but true in any order of the perturbation  theory. One can check it
using Feynman parameters, and general properties of Symanzik polynomials
\cite{Bogner:2010kv}. Taking this observation into account, we replace the last factor in
Eq.~(\ref{polar}) as follows:
\begin{equation}
    \label{replace}
    \theta(-k^2_\perp)\to\theta(-k^2_\perp)-\theta(k^2_\parallel).
\end{equation}
We stress that this substitution of the measure does not change the right hand side of Eq. (\ref{SD}). Furthermore,
the difference of theta functions can be represented as follows:
\begin{equation}
    \label{differnce}
    \theta(-k^2_\perp)-\theta(k^2_\parallel)=
    -\frac{k^2_\parallel}{|k^2_\parallel|}
    \theta(\nu)\theta(1-\nu),
\end{equation}
where we introduced a new variable $\nu$, which we will call
the \textit{inclination} of the fermion with momentum $k$:
\begin{equation}
\label{inclination}
\nu\equiv\frac{k_\parallel^2}{k^2}.
\end{equation}
This variable is in one-to-one correspondence with $k_\perp^2$:
\begin{equation}
    \label{k-perp-nu}
    k_\perp^2=k_\parallel^2(1/\nu-1).
\end{equation}

 Notice that when $k^2_\parallel>0$, and
$0<\nu<1$, the components of $k_\perp$ are purely imaginary. We say that
using inclination leads one to consider \textit{doubly virtual} particles, by which we
mean particles not only away from the mass shell, but also with imaginary momentum
components. 

The last step in the transformation of the integration measure which we need
is to replace the variable $k^2_\perp$ with the inclination variable:
\begin{equation}
    \label{change}
    dk^2_\perp=|k_\parallel^2|d\nu/\nu^2.
\end{equation}

We conclude that Schwinger-Dyson equation (\ref{SD}) can be rewritten as follows:
\begin{widetext}
\begin{equation}
    \label{SDi}
    \Gamma_\mu(p,p')=\gamma_\mu+ 
\frac{\alpha}{2\pi}\int_0^{2\pi}\frac{d\phi}{2\pi}
\int_0^1d\nu\int\frac{d^2k_\parallel}{2\pi}
\frac{\Gamma_{\mu\lambda}(k,p';p-k)}
{(k_\parallel-\nu p)^2+p^2\nu(1-\nu)+i\epsilon}
\big(i(p-k)^2D^{\lambda\rho}(p-k)\big)
   \slashed{k}\gamma_\rho.
\end{equation}
\end{widetext}
It is tacitly assumed in this formula that the transverse
components of $k$ are expressed in terms of the integration
variables, for which one uses Eqs. (\ref{components}) and
(\ref{k-perp-nu}). As mentioned above, integration over the
inclination implies integration over the momentum of the doubly virtual particle.

The form of Eq.~(\ref{SDi}) explains the origin of the name ``inclination'' for
the variable $\nu$: at $p^2\to 0$ a vicinity of the point $k_\parallel=\nu p$ gives a dominant 
contribution to the integral. So, one can say that $\nu$ is a fraction of the original
momentum which the fermion is inclined to keep after emitting a photon. 

In the next Section we expound on this, and see how restricting the integration to a vicinity
of the point $k_\parallel= \nu p$ allows one to truncate the Schwinger-Dyson equation (\ref{SDi}) to a
closed equation for the vertex.

\section{Truncation of the Schwinger-Dyson Equation}

First we point out that if $k_\parallel=\nu p$, Eq. (\ref{k-perp-nu}) 
implies that $k_\perp\to 0$ when $p^2\to 0$. Therefore, in
the leading approximation $k_\perp$ can be set
to zero in the right hand side of Eq. (\ref{SDi}) and the integration in 
the angle $\phi$ removed. Second, the factor $\slashed{k}\gamma_\rho$ can
be replaced as follows:
\begin{equation}
\label{k-gamma}
\slashed{k}\gamma_\rho\to 2\nu(p-k)_\rho/(1-\nu),
\end{equation}
because $\slashed{k}\gamma_\rho=2\nu p_\rho-\gamma_\rho\slashed{p}$ at 
$k=\nu p$, and
$\slashed{p}\approx 0$ in the rightest position.

Next simplification which one can make near the point $k_\parallel=\nu p$
is available if the dimensional regularization unit of mass $\mu^2$ normalizing the
renormalized fine structure constant is taken to be equal to $-p^2$. In this case
the vacuum polarization involved in the photon propagator may be set to zero, because 
it depends on $\log(-p^2/\mu^2)=0.$ This means that photon propagator in 
Eq.~(\ref{SDi}) can be replaced with the free one.

The last simplification which we make is removing the longitudinal part of the free photon
propagator. We can do it in the leading approximation, because the longitudinal part
does not give a contribution logarithmic in $-p^2.$

After these replacements, Schwinger-Dyson equation (\ref{SDi}) takes the following form:
\begin{widetext}
\begin{equation}
    \label{SDir}
    \Gamma_\mu(p,p')\approx\gamma_\mu+ 
\frac{\alpha}{2\pi}
\int_0^1\frac{\nu d\nu}{1-\nu}\int_V\frac{d^2k_\parallel}{\pi}
\frac{(p-k_\parallel)^\rho\Gamma_{\mu\rho}(k_\parallel,p';p-k_\parallel)}
{(k_\parallel-\nu p)^2+p^2\nu(1-\nu)+i\epsilon}
.
\end{equation}
\end{widetext}
Notice that the integration in $k_\parallel$ is now restricted to
a vicinity $V$ of the point $k_\parallel=\nu p.$ The approximate equality
means that only the leading contribution at $p^2\to 0$ is kept in the
right hand side.

Now we can use Ward-Takahashi identity for the four-point amplitude featuring
Eq. (\ref{SDir}). For this particular connected amplitude it reads
\begin{widetext}
\begin{equation}
    \label{WT}
    (p-k_\parallel)^\rho\Gamma_{\mu\rho}(k_\parallel,p';p-k_\parallel)
    =-i\big[\Gamma_\mu(p,p')\frac{\slashed{p}\slashed{k}_\parallel}{p^2}
    -S^{-1}(p')S\big(p'-(p-k_\parallel)\big)\Gamma_\mu\big(k,p'-(p-k_\parallel)\big)\big],
\end{equation}
\end{widetext}
where $S(l)$ denotes the full fermion propagator depending on momentum $l.$ In the leading
approximation, the full propagators can be replaced with the free propagators,
because the dependence on the normalization point cancels in the ratio of the propagators.
Finally, setting $k_\parallel=\nu p$ and neglecting $p^2$ where possible, we obtain 
after a simple rearrangement of the terms the form of the Ward-Takahashi identity we will use:
\begin{widetext}

\begin{equation}
    \label{WTf}
    (p-k_\parallel)^\rho\Gamma_{\mu\rho}(k_\parallel,p';p-k_\parallel)\approx
    -i\big[\Gamma_\mu(p,p')\nu-\Gamma_\mu\big(k_\parallel,p'-p(1-\nu)\big)
    +\frac{(1-\nu)\slashed{p}\slashed{p'}}{(2pp')(y+1-\nu)}
    \Gamma_\mu\big(k_\parallel,p'-p(1-\nu)\big)\big].
\end{equation}
\end{widetext}
We recall that $y=-p'^2/(2pp').$
Notice that $k_\parallel$ is not replaced with $\nu p$
in the first argument of
$\Gamma_\mu.$ This is because at $p^2=0$ such a replacement would give
infinity. 

Substituting Eq.~(\ref{WTf}) into Eq.~(\ref{SDir}) gives the truncation
of Schwinger-Dyson equation we aimed at:
\begin{widetext}

\begin{align}
    \label{truncation}
    &\Gamma_\mu(p,p')\approx\gamma_\mu-\frac{\alpha}{2\pi}
\int_0^1\frac{\nu d\nu}{1-\nu}\int_V\frac{id^2k_\parallel/\pi}
{(k_\parallel-\nu p)^2+p^2\nu(1-\nu)+i\epsilon}\times\nonumber\\
&\times
\big[\Gamma_\mu(p,p')\nu-\Gamma_\mu\big(k_\parallel,p'-p(1-\nu)\big)+\frac{(1-\nu)\slashed{p}\slashed{p'}}{(2pp')(y+1-\nu)}
    \Gamma_\mu\big(k_\parallel,p'-p(1-\nu)\big)\big]
\end{align}
\end{widetext}
This equation can be used to generate iteratively the expansion in $\alpha$ of the vertex $\Gamma_\mu$
starting from the initial term $\gamma_\mu.$ This is the subject of the next section.

\section{The Form Factor of the Skewed Sudakov Regime}

To compute the first correction in the perturbative expansion of the vertex, one takes
the second term in the right hand side of Eq.~(\ref{truncation}) and makes in it the substitution
$\Gamma_\mu\to\gamma_\mu$ regardless of the arguments of $\Gamma_\mu$. 
One keeps in the integrals only the leading terms at $p^2\to 0.$
The outcome of this exercise is as follows:
\begin{widetext}
\begin{equation}
    \label{first}
    \Gamma^{(1)}_\mu(p,p')\approx\gamma_\mu\big[t\int_0^1d\nu\big(\nu-\frac{\nu z}{1-\nu z}\big)\big]
    +\frac{p_\mu\slashed{p'}}{pp'}\big[t\int_0^1d\nu\frac{\nu z}{1-\nu z}\big],
\end{equation}
\end{widetext}
where the first correction is in the left hand side, and the variables $t, z$ were defined in Eq. (\ref{tz}).
A relation useful in checking this reads
\begin{equation}
    \label{useful0}
\frac{\alpha}{2\pi}\int_V\frac{id^2k_\parallel/\pi}
{(k_\parallel-\nu p)^2+p^2\nu(1-\nu)+i\epsilon}\approx t.   
\end{equation}

With the first correction known, one can use it to obtain the second one, and so
on. A minor complication is that the integration in $k_\parallel$ requires
the knowledge of $\Gamma_\mu\big(k_\parallel,p'-p(1-\nu)\big)$ at small $k^2_\parallel$
of any sign, and Eq.~(\ref{first}) gives it only for $k^2_\parallel<0$. As follows from
the derivation of Eq.~(\ref{first}), it can be extended to positive small virtuality
of the incoming fermion by making the replacement $\log(x)\to\log(|x|)$ in
Eq.~(\ref{tz}). Apart of this clarification, one also needs in this computation
the following 
generalization of Eq.~(\ref{useful0}):
\begin{equation}
    \label{usefuln}
\frac{\alpha^{(n+1)}}{(2\pi)^{(n+1)} n!}\int_V\frac{\log^n\big(|2pp'/k_\parallel^2|\big)id^2k_\parallel/\pi}
{(k_\parallel-\nu p)^2+p^2\nu(1-\nu)+i\epsilon}\approx \frac{t^{(n+1)}}{(n+1)!},   
\end{equation}
where $n$ is any integer.

The outcome of the approximate computation of the perturbative corrections outlined above
can be summarized as follows. First, no new tensor structures appear in the vertex apart of the
ones already present in the leading correction of Eq. (\ref{first}):

\begin{equation}
    \label{F-Phi}
    \Gamma_\mu(p,p')\approx\gamma_\mu F(t,z)
    +\frac{p_\mu\slashed{p'}}{pp'}\Phi(t,z).
\end{equation}
Second, the form factors satisfy the following evolution equations:
\begin{widetext}
\begin{align}
    \label{F-e}
    &\frac{\partial F(t,z)}{\partial t}=\frac{1}{2}F(t,z)-\int_0^1d\nu
    \Big[\frac{\nu\big(F(t,z)-F(t,\nu z)\big)}{1-\nu}+
    \frac{\nu z F(t,\nu z)}{1-\nu z}\Big],\,F(0,z)=1.\\
    \label{Phi-e}
    &\frac{\partial \Phi(t,z)}{\partial t}=\frac{1}{2}\Phi(t,z)-\int_0^1d\nu
    \Big[\frac{\nu\big(\Phi(t,z)-\Phi(t,\nu z)\big)}{1-\nu}-
    \frac{\nu z F(t,\nu z)}{1-\nu z}\Big],\,\Phi(0,z)=0.
\end{align}
\end{widetext}

Next we notice that the sum of these equations yields a simpler equation, which can be solved:
$$F(t,z)+\Phi(t,z)=e^{t/2}$$. We conclude that the vertex is approximated with a single form
factor, Eq.~(\ref{sS}),
which satisfies the evolution equation (\ref{F-e}).

\section{Solving Evolution Equation for the Form Factor}

Eq.~(\ref{F-e}) is a linear evolution equation of the form
$\dot{F}=\mathcal{O}F$, where the linear operator
$\mathcal{O}$ acts on a function of $z$:
\begin{equation}
    \label{O}
    \big(\mathcal{O}\chi\big)(z)=\frac{\chi(z)}{2}-\int_0^1 d\nu
    \Big[\frac{\nu\big(\chi(z)-\chi(\nu z)\big)}{1-\nu}+
    \frac{\nu z \chi(\nu z)}{1-\nu z}\Big].
\end{equation}

One explicitly checks that $\chi_n(z)=z^n(1-z)$ are eigenfunctions of
the operator $\mathcal{O}$:
\begin{equation}
    \label{spectrum}
    \mathcal{O}\chi_n = \big(\frac{3}{2}-H_{n+1}\big)\chi_n,
\end{equation}
and the initial condition for the form factor can be expanded in terms of these
eigenfunctions:
\begin{equation}
    \label{ini}
    F(0,z)=\sum_{n=0}^\infty \chi_n(z).
\end{equation}
This implies Eq. (\ref{F}).

Next we derive the representation (\ref{L}) starting from Eq.~(\ref{F}). According to it, expansion
of $(e^{-3t/2}F(t,z)-1)z\equiv\bar{F}(t,z)$ in powers of $t$ is as follows:
\begin{equation}
    \label{S}
    S_k(z)=\frac{1}{k!}\sum_{n=1}^\infty H^k_n z^n(1-z), \, 
    \bar{F}(t,z)=\sum_{k=1}^\infty (-t)^kS_k(z)
\end{equation}
Our task is to express $S_k(z)$ in terms of 
multiple polylogarithms. 

To complete the formulation of the task we give a definition of the multiple polylogarithms
in terms of linear operators acting on a function of a single variable, which is equivalent to
the standard definition in terms of the iterated integrals \cite{Waldschmidt}. First we define 
two operators used in the definition:
\begin{align}
    \label{AE}
    &\big(A\phi\big)(z)=\int_0^z\frac{dz'}{z'}\phi(z'),\nonumber\\
    &\big(E\phi\big)(z)=\int_0^z\frac{dz'}{1-z'}\phi(z').
\end{align}
Here $\phi(z)$ is a function on which the operators $A$ and $E$ act. 

Next we define an operator labelled with a string of integers $\underline{s}=(s_1,\dots,s_d), s_i>0$:
\begin{equation}
    \label{L-s}
    L_{\underline{s}}=A^{s_1-1}EA^{s_2-1}E\dots A^{s_d-1}E,
\end{equation}
where $d$ is the depth of the string $\underline{s}.$

With these notations a multiple polylogarithm is defined as follows:
\begin{equation}
    \label{MPL}
    Li_{\underline{s}}(z)=(L_{\underline{s}}u)(z),
\end{equation}
where the operator $L_{\underline{s}}$ acts on  the function $u(z)=1.$

We now return to transforming $S_k(z)$. Expansion of $S_k(z)$ in powers
of $z$ reads
\begin{equation}
    \label{S-exp}
    S_k(z)=\frac{z}{k!}+\sum_{n=2}^\infty z^n
    \frac{H_n^k-H^k_{n-1}}{k!}.
\end{equation}
Since $H_n=H_{n-1}+1/n$, one can expand its power in Eq.~(\ref{S-exp}), and
cancel the term $H^k_{n-1}$ in the numerator. A straightforward 
regrouping of terms transforms thus Eq. (\ref{S-exp}) as follows:
\begin{equation}
    \label{S-exp-1}
    S_k(z)=\frac{1}{k!}\sum_{n=1}^\infty\frac{z^n}{n^k}+
    \sum_{n=2}^\infty z^n\sum_{l=1}^{k-1}\frac{H^l_{n-1}}{n^{k-l}l!(k-l)!},
\end{equation}
where the second sum vanishes for $k=1$.

Next we notice that $Ax^n=x^n/n$, and one can write the above expansion using 
the operators $A$ and $E$:
\begin{equation}
    \label{S-last}
    S_k(z)=\frac{1}{k!}\big(A^{(k-1)}Eu\big)(z)+\sum_{l=1}^{k-1}\frac{1}{l!}A^lES_{k-l}(z).
\end{equation}
This gives a recursive definition of $S_k(z)$ in terms of the operators $A, E$ and function $u$
starting with $S_1(z)=(Eu)(z)$.

At last, one proves by induction that this recursion yields the following expression for
$S_k(z)$
\begin{equation}
    \label{finale}
    S_k(z)=\sum_{|\underline{s}|=k}\frac{1}{\underline{s}!}(L_{\underline{s}}u)(z),
\end{equation}
which implies Eq.~(\ref{L}).

\section{Conclusions}

In this paper we have given a generalization (\ref{sS}) of the Sudakov approximation (\ref{Sudakov})
valid in a wider kinematic range. Considering the abundant literature 
(see, e.g., \cite{Sen:1981sd}, \cite{Korchemsky:1988hd}, \cite{Larkoski:2015lea})  derived in
various ways from \cite{Sudakov:1954sw} one may envisage a scientific program 
trying to give a skewed version to any result descending from Ref.~\cite{Sudakov:1954sw}.
 
Our subjective choice for the sequence of these possible generalizations is as follows:
First, one may try to study the skewed asymptotic for non-abelian gauge theories \cite{Sen:1981sd}, second,
the subleading corrections \cite{Korchemsky:1988hd}, third, the phenomenology \cite{Larkoski:2015lea}.

The idea of this paper belongs to Lev Lipatov, who passed away before the first draft of the 
paper was ready. His profound influence contributed substantially to our studies and, in particular, to this paper.
V.~K. and G.~P. thank V.~S.~Fadin for reading an early draft and for  fruitful discussions. G.~P. thanks
CERN Theory Department for the kind hospitality.


\begin{thebibliography}{9}
\bibitem{Sudakov:1954sw} 
  V.~V.~Sudakov,
  Sov.\ Phys.\ JETP {\bf 3}, 65 (1956)
  [Zh.\ Eksp.\ Teor.\ Fiz.\  {\bf 30}, 87 (1956)].
  
  \bibitem{Waldschmidt}
  M. Waldschmidt, Multiple Polylogarithms: An Introduction, in 
  ``Number Theory and Discrete Mathematics,'' Editors:  A. K. Agarwal et al., 
  Birkhäuser Basel 2002, DOI 10.1007/978-3-0348-8223-1
  
\bibitem{Weinzierl:2007cx} 
  S.~Weinzierl,
  IRMA Lect.\ Math.\ Theor.\ Phys.\  {\bf 15}, 247 (2009)
  doi:10.4171/073-1/8
  [arXiv:0705.0900 [hep-ph]].
  
\bibitem{Bogner:2012dn} 
  C.~Bogner and F.~Brown,
  PoS LL {\bf 2012}, 053 (2012)
  doi:10.22323/1.151.0053
  [arXiv:1209.6524 [hep-ph]].
  
\bibitem{Bogner:2010kv} 
  C.~Bogner and S.~Weinzierl,
  Int.\ J.\ Mod.\ Phys.\ A {\bf 25}, 2585 (2010)
  doi:10.1142/S0217751X10049438
  [arXiv:1002.3458 [hep-ph]].
  
\bibitem{Sen:1981sd} 
  A.~Sen,
  Phys.\ Rev.\ D {\bf 24}, 3281 (1981).
  doi:10.1103/PhysRevD.24.3281

\bibitem{Korchemsky:1988hd} 
  G.~P.~Korchemsky,
  Phys.\ Lett.\ B {\bf 220}, 629 (1989).
  doi:10.1016/0370-2693(89)90799-5


\bibitem{Larkoski:2015lea} 
  A.~J.~Larkoski, S.~Marzani and J.~Thaler,
  Phys.\ Rev.\ D {\bf 91}, no. 11, 111501 (2015)
  doi:10.1103/PhysRevD.91.111501
  [arXiv:1502.01719 [hep-ph]].
  
  
\end{thebibliography}
\end{document}